# Galaxy Number Counts

## Ana Campos

*Physics Dept., Univ. of Durham, South Road, Durham DH1 3LE.*

9 October 1995

## ABSTRACT

In these lecture notes I re-visit the problem of the deep number counts as a viable method to (1) study the evolution of the galaxy population in the past and (2) determine the value of the deceleration parameter $q_0$. After a brief description of the basic concepts needed to model the counts, we show that simple luminosity evolution models which incorporate dust are able to provide reasonable fits to both the deep counts and the redshift distribution of faint field galaxies if $q_0 = 0.05$. In order to fit the counts in a *closed* ($q_0 = 0.5$) Universe, we include a population of fading dwarfs, more numerous in the past, based on current ideas to understand the nature of dwarf galaxies.

## 1 INTRODUCTION

Galaxy number counts have been widely-used in Cosmology in an attempt to estimate the value of the deceleration parameter $q_0$ (throughout this paper I will assume $\Lambda = 0$). Whereas most observational approaches are based on the direct estimation of the density of matter in the Universe (eg. superclusters mass/volume), number counts deal with the pure geometric effect of varying $q_0$.

The idea is fairly simple. Let us imagine that we take a very deep wide-field image at any place on the sky, counting afterwards the number of galaxies per unit area and magnitude bin ($N(m)$). The value of $q_0$ can be inferred by means of the comparison of the *observed* $N(m)$ with the predictions from different $q_0$ models. In the most basic models, usually called no-evolution models, two assumptions are made: (1) the luminosity of the galaxies remains constant on time, (2) the number of galaxies per unite volume is conserved after they were formed.

Since the first papers on number counts (Brown & Tinsley 1974; Bruzual & Kron 1980; Shanks et al. 1984; King & Ellis 1985; Koo 1986) it has been well known that these simple no-evolution models are unable to fit the counts, no matter what value of $q_0$ is used. There is an *excess* of galaxies at faint magnitudes (B fainter than $\sim 22-23$ depending on the normalization), which is much larger when the counts are made using blue photometric passbands. Thus, this problem is usually referred to as the excess of faint *blue* galaxies.

In order to reconcile observations with model predictions it is neccesary to account for evolution in the galaxy population. The luminosity of the galaxies is expected to vary with time, as stars are evolving. In particular we expect that galaxies were brighter at earlier times, closer to the epoch at which they were formed. After the first episodes of star formation the birth rate of stars is expected to decrease (either because the interstellar gas is exhausted or because it is blown out of the galaxy), like it seems to be the case in early type galaxies, or to remain approximately constant on time, as we observe in late type galaxies. If galaxies were brighter in the past then for a given apparent magnitude they are observable at greater redshifts, and so for the same angular area we survey a larger volume (Tinsley 1980; Guiderdoni & Rocca-Volmerange 1990; Metcalfe et al. 1991).

It has also been suggested (Guiderdoni & Rocca-Volmerange 1991; Broadhurst et al. 1992) that not only the luminosity but also the number density of galaxies could vary with time. This variation is, for example, expected in hierarchical models of galaxy formation where galaxy sub-units merge to build-up the present day population. In these models galaxies are assumed to be more numerous in the past, and so they predict larger numbers of galaxies to be seen at faint magnitudes.

In this paper we re-visit the modelling of the number counts by means of luminosity evolution models which incorporate dust. In Section 2 we briefly summaryze the basic concepts to model the predictions in the no-evolution case. The effects of the luminosity evolution (and the dust) in the model predictions are analysed in Section 3. Section 4 is devoted to the study of the redshift distributions of faint field galaxies, as an independent test to the number counts models. In Section 5 we show that in order to fit the counts in a $q_0 = 0.5$ Universe in the framework of luminosity evolution models, we must allow for luminosity-dependent luminosity evolution. We also show that the introduction of a population of fading dwarfs allows to fit the counts. Finally, the conclusions are presented in Section 6.

## 2 MODELLING THE NUMBER COUNTS

The number of galaxies with apparent magnitude $[m, m+dm]$ located in the redshift shell $[z, z+dz]$, which contributes to the counts per steradian can be computed as follows:

$$d^2 A(m,z) = \Phi(M)(1+z)^3 \frac{dV}{dz} dm dz \qquad (1)$$

**Table 1.** Luminosity Functions

| Galaxy type | $\phi^\star$(%) | $\alpha$ | $M^\star$ |
|---|---|---|---|
| Early | 57 | -0.7 | -19.6 |
| Intermediate | 26 | -1.1 | -19.9 |
| Late | 17 | -1.5 | -20.0 |

where $\Phi(M)$ is the luminosity function (LF; it can be estimated using nearby redshift surveys) and $M$ is the absolute magnitude of a galaxy observed to have an apparent magnitude $m$ when it is located at a redshift $z$. $dV/dz$ is the volume element per steradian of the redshift shell $[z, z + dz]$, which in the particular case of Friedman models depends on the values of $q_0$ and $H_0$ (see eg. Weinberg 1972).

The relation between the absolute magnitude of a galaxy with its apparent magnitude at any $z$ is given by:

$$m = 5\log(D_L) + M + K(z) \qquad (2)$$

where $D_L$ is the luminosity distance, which again in Friedman models depends on the values of $q_0$ and $H_0$. $K(z)$ is the $K$-correction, which accounts for the red-shifting of the spectral light due to the Doppler effect. For any given photometric passband we observe different wavelength ranges (in the rest-frame of the galaxy) depending on the redshift at which galaxies are located, and so we have to do the appropiate correction because the spectra of galaxies are not flat. For the no-evolution models, the $K$-corrections are computed using spectra of nearby galaxies.

The number of galaxies per steradian and magnitude bin can be obtained by integrating $d^2A$ from $z = 0$ up to the redshift at which galaxies were formed $z_{for}$,

$$N(m)dm = \int_{z=0}^{z_{for}} d^2A(m, z) \qquad (3)$$

In what follows we will use the Schechter parametrization of the LF, which has the form:

$$\begin{aligned}\Phi(M) &= 0.92\Phi^\star \exp[-0.92(\alpha + 1)(M - M^\star)\\ &\quad - \exp[-0.92(M - M^\star)]]dM\end{aligned} \qquad (4)$$

where $\alpha$ is the slope of the function at its faint end and $M^\star$ is a characteristic magnitude at which the *knee* of the function occurs. $\phi^\star$ is the normalization, i.e. the number density of galaxies. Notice that both $\phi^\star$ and $M^\star$ depend on $H_0$. The dependence of $\phi^\star$ and $dV/dz$ on $H_0$ cancels out in equation (1). On the other hand, the dependence of $D_L$ on $H_0$ cancels with that of $M^\star$. Therefore the number counts only depends on the value of $q_0$ (and of course $\Lambda$, in the case it had a value different than zero).

In Figure 1 we show the number counts in the B-band together with predictions from no-evolution models. The data is a compilation from the literature: Kron 1978; Koo 1986; Jarvis & Tyson 1981; Couch & Newell 1984; Infante et al. 1986; Tyson 1988; Jones et al. 1990; Lilly et al. 1991; Metcalfe et al. 1991, 1995. To compute the predictions we have assumed three different galaxy types, each one with different $K$-corrections and LF. The LFs (see Table 1) have been obtained from the Durham-redshift survey (Shanks et al. 1984; see also Metcalfe et al. 1991). As we already mentioned in the introduction, the no-evolution models are unable to fit the counts, even for low-$q_0$ values for which the volume at high-$z$ is larger.

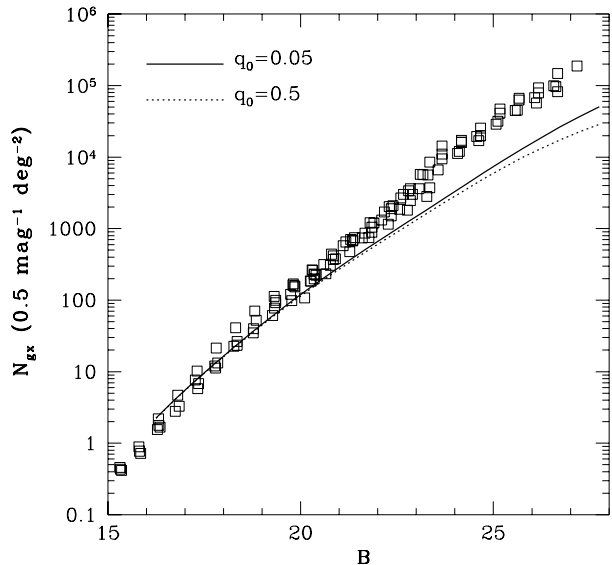

**Figure 1.** The number counts in the B band. The data has been taken from the literature (see text for references). Also we plot predictions from no-evolution models.

## 3   LUMINOSITY EVOLUTION MODELS

The luminosity of the galaxies is expected to evolve with time. A simple method to include luminosity evolution in the models is by adding an extra-term in the right-hand side of equation (2) as follows:

$$m(z) = 5\log(D_L) + M(z = 0) + K(z) + e(z) \qquad (5)$$

where $e(z)$ is the evolution correction, which measures the variation of the luminosity (in units of magnitudes) from redshift $z$ to $z = 0$. In this equation we now relate the apparent magnitude of a galaxy located at a redshift $z$ with the absolute magnitude that the same galaxy will have at $z = 0$ (this is because in equation (1) we will make use of the LF estimated using nearby surveys, i.e. at $z = 0$).

The $K$- and $e$- corrections ($K + e$ in what follows) are computed by means of spectrophotometric models of stellar populations (eg. Tinsley 1972; Bruzual 1981; Arimoto & Yoshii 1986, 1987; Bruzual & Charlot 1993). These models are based on stellar libraries which provide information on the stellar spectra (or colours) at any position in the HR diagram, i.e. at any stage of the stellar evolution. We can model the spectral evolution of a galaxy by assuming a redshift of formation ($z_{for}$), an Initial Mass Function (IMF) for the stars and a Star Formation Rate (SFR). To check whether the chosen model parameters ($z_{for}$, IMF and SFR) are appropiate one compares the synthetic spectrum at $z = 0$ with observed spectra of nearby galaxies. Finally, once we have the spectra of the model galaxy at any redshift, it is straighforward to compute its magnitude (at any $z$) just simply by convolving the (red-shifted) synthetic spectrum with the photometric passband. The difference between the magnitude at $z$ and at $z = 0$ will give the $K + e$ corrections.

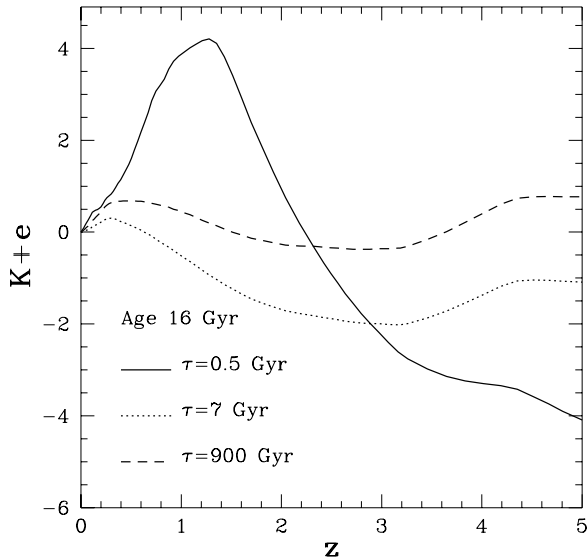

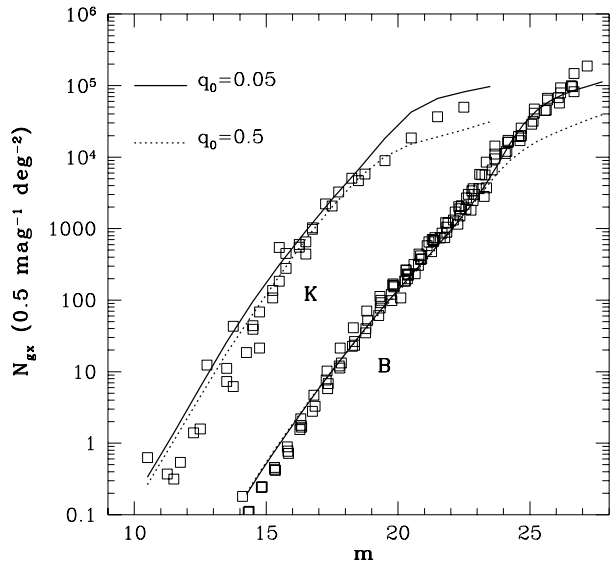

**Figure 2.** K-correction plus evolution for three types of galaxies. The age of the galaxies at $z = 0$ is 16 Gyr., which corresponds to a redshift of formation of $z = 6$ for $q_0 = 0.05$ and $H_0 = 50$ km s$^{-1}$ Mpc$^{-1}$.

**Figure 3.** The number counts in the B and K bands, and predictions from luminosity evolution models with dust.

We notice that when the luminosity evolution is included to model the counts, the predictions depend ($\sim$ weakly) on the value of $H_0$. The reason is found in the relation between age and redshift (which depends on $H_0$) to trace the galaxy evolution.

In Figure 2 we have plotted the $K + e$ corrections for the B-band for three model-galaxy types: early (E/S0), intermediate (Sa/Sb/Sc) and late (Sd/Irr). The age of the galaxies is 16 Gyr ($z_{for} = 6$ for $q_0 = 0.05$ and $H_0 = 50$ km s$^{-1}$ Mpc$^{-1}$) in all cases, and we have used the Salpeter IMF. The SFR is assumed to decay exponentially, with different e-folding times: 0.5, 7 and 900 Gyr for the early, intermediate and late types respectively. To compute the $K + e$ corrections we have used the spectrophotometric models of Bruzual & Charlot (1993). As can be seen in the Figure, the $K + e$ corrections for late-type galaxies are very small, because their SFR is nearly constant but also because their spectra are quite flat. On the contrary, for early type galaxies the $K + e$ corrections are very large. The observed luminosity in the B-band for $z > 0$ is smaller than for $z = 0$, up to a redshift of $z \sim 2$. The reason is that the UV-light enters into the B-band, and early-type galaxies are not strong emitter in the UV. However for redshifts $z > 2$ the galaxies become more and more luminous, as we move closer to the epoch were the bulk of stars was formed.

In Figure 3 we plot the observed number counts in the B and K bands together with predictions from $q_0 = 0.05$ and $q_0 = 0.5$ luminosity evolution models. The data in the K-band has been taken from Mobasher et al. 1986; Glazebrook et al. 1993; Jenkins & Reid 1991 and Gardner et al. 1993. To compute the models we have assumed three dif-ferent galaxy types as in Figure 2, and the LFs shown in Table 1. For the $q_0 = 0.5$ model we assume 12.7 Gyr, which corresponds to $z_{for} = 10$, rather than 16 Gyr for the age of the galaxies (the reason is that in a $q_0 = 0.5$ and $H_0 = 50$ model the age of the Universe is only 13.1 Gyr). The models have been normalized to fit the number counts in the B-band at $B = 17.5$, which corresponds to an overall normalization of $\phi^* = 1.4 \times 10^{-2}$ h$^{-1}$ Mpc$^{-3}$. The counts in the K-band were derived using the luminosity function in the B-band by means of the B$-$K colours computed for each model galaxy.

To model the number counts we have also taken into account the presence of dust in galaxies. The effect of dust, neglected in many works on number counts, is proving to be very important (Wang 1991; Campos & Shanks 1995). The dust absorbs the light coming from the stars, re-radiating it in the IR. In general the effect of dust is much more severe as the wavelength shortens. This fact is reflected in the extinction law, which gives the dependence of the optical depth ($\tau$) on wavelength. A very rough analytical approximation to the extinction law can be written in the form $\tau \propto \lambda^{-n}$, with $n \sim 1 - 2$ (Draine & Lee 1984). The extinction properties of dust particles are still poorly known, both empirically or *via* modelling. As a further complication, it is known that the dust obscuration depends on how stars and dust are mixed together, so that the extinction law can differ not only from galaxy type to galaxy type but even from galaxy to galaxy. Besides, little is known about the relative importance of absorption to scattering, as well as on the evolution of dust in galaxies. Dust is formed during the post-main sequence evolution of massive stars, is therefore expected to appear in the galaxies soon after the formation of the first stars.

In the number counts predictions presented here we have used a very simple model for dust, which is the same for all galaxy types and at any time (i.e. we do not model dust evolution). The extinction law is taken to be $\propto \lambda^{-2}$, and the optical depth of an $L_*$ galaxy is assumed to be $\tau_* = 0.3$ in the B-band (which, in a standard model, corresponds to an extinction of $A_B \sim 0.15$ mag at $z = 0$). We also assume that the optical depth depends on the galaxy luminosity as $\tau \propto L_{z=0}^{0.5}$, to account for the fact that low luminosity nearby galaxies seem to be less obscured by dust. In any case it is worth mentioning that this assumption does not have strong effect on the predictions (Campos & Shanks 1995). As we will comment later, the counts are dominated by $L_*$ galaxies, except at the faint limits where the presence of the intrinsically faint objects becomes very important.

The importance of dust in the number counts models is mainly due to the combination of spectral red-shifting due to the Doppler effect and the shape of the extinction law. The effect of dust is more important as the redshift of the galaxy increases, because we observe light emitted at shorter wavelengths where the extinction is larger. Thus, the extinction by dust has the effect of *dimming* the brightening of the luminosity at high redshifts. In fact, luminosity evolution models in which dust is not included overpredict the counts at faint limits.

As can be seen in Figure 3, the $q_0 = 0.05$ model provides a very good fit to the data in $\sim 10$ magnitudes interval range ($B \sim 17 - 27$). The $q_0 = 0.5$ model fails to fit the counts fainter than $B \sim 24$, as it predicts fewer galaxies than observed. With respect to the K band, the $q_0 = 0.05$ model slightly overpredicts the counts at both the bright and the faint end. The $q_0 = 0.5$ model gives better fit to the data down to $K \sim 21$, but underpredicts the counts fainter than that. The discrepancies between models and data in the K-band are not very large, and so we regard them as due to uncertainties in the spectrophotometric models when computing the B−K colour of the model galaxies.

## 4    THE REDSHIFT DISTRIBUTIONS OF FAINT FIELD GALAXIES

In order to constrain the number counts models we can make use of the redshift distribution $N(z)$ of faint field galaxies as an independent test. To compute the $N(z)$ predictions for galaxies selected in the magnitude range $[m_1, m_2]$ we just have to integrate:

$$N(z)dz = \int_{m_1}^{m_2} d^2 A(m, z) \qquad (6)$$

In Figures 4 and 5 we plot the $N(z)$ distributions of galaxies selected in the B-band to have an apparent magnitude of $21 <$B$< 22.5$ (data taken from Colless et al. 1990, 1993) and $23 <$B$< 24$ (data taken from Glazebrook et al. 1995 and Campos et al. 1995) respectively. In the same Figures we also plot predictions from the luminosity evolution models (with dust) presented in the previous section. As can be seen, the models fit fairly well the data for the shallow sample, although for the deep one they predict a *high redshift tail* which is not observed. The $N(z)$ distribution is quite close to the predictions from no-evolution models, also shown in the Figures. In fact this came as a surprise, as the no-evolution models are unable to fit the counts.

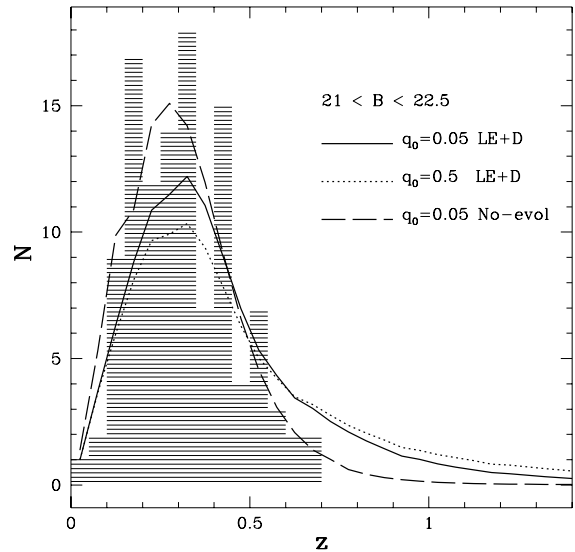

**Figure 4.** The redshift distribution of galaxies of $21 < B < 22.5$ and predictions from no-evolution and luminosity evolution models with dust (LE+D).

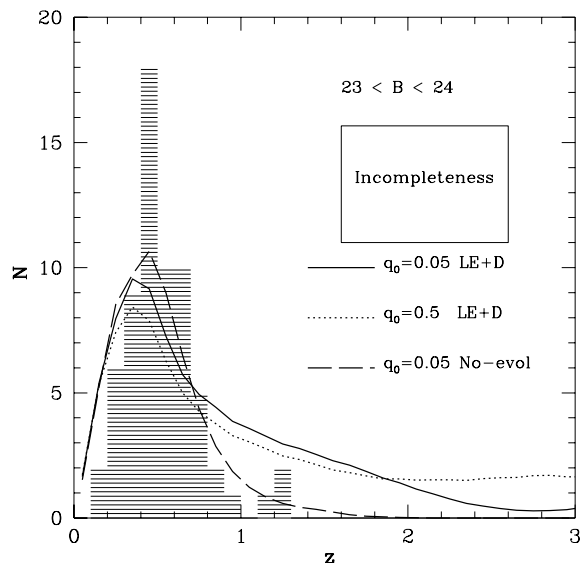

**Figure 5.** The same as Figure 4 but for $23 < B < 24$ galaxies.

The discrepancies between data and predictions from luminosity evolution models might be due to an observational bias rather than a *failure* of the models (Campos et al. 1995). The deep sample is incomplete, i.e. the redshift of ∼ 40% of the objects could not be identified as no spectral features were detected. If these galaxies turned out to be at high−z then the models would fit the observations rather well. There is a very simple reason which allows us to suggest that this might actually be the case. If a galaxy was at a redshift higher than z ∼ 1, then the (quite often strong) [OII]λ3727 emission line would be red-shifted outside the optical spectral window (usually running from ∼ 4000 to 8000 Å). At shorter wavelengths there are of course other spectral features, such as C, Fe or Mg lines, but these are usually very weak and so difficult to detect in these faint galaxies. We therefore expect that there is a bias *against* the redshift-identification of high-z objects. A further indication pointing out that this might actually be the case is found in the colour distribution of the objects. The models predict that this evolved high-z population is *bluer*, and, as shown by Campos et al. (1995), the unidentified galaxies are bluer on average than the galaxies with measured redshifts.

The importance of dust to model the counts is further evidenced in the $N(z)$ predictions. The same models presented here but without dust (Campos & Shanks 1995) predict a much larger redshift tail, totally incompatible with the data even in the case that the unidentified galaxies were found to be at high-z.

## 5  COULD WE FIT THE COUNTS IN A $Q_0 = 0.5$ UNIVERSE MODEL?

We have shown in the previous sections that simple luminosity evolution models which incorporate dust are able to provide reasonable fits to the counts and the redshift distributions, but only when $q_0 = 0.05$. As shown in Figure 3, the $q_0 = 0.5$ model underpredicts the counts fainter than $B \sim 24$.

In general, at *any* apparent magnitude range the counts are dominated by $L_*$ galaxies. The contribution of galaxies brighter than $L_*$, even if they are placed at higher-z and so we survey a larger volume for the same apparent magnitude, is smaller simply because they are less abundant. The number density of galaxies fainter than $L_*$ is larger, but for the same apparent magnitude range they are closer and so in a smaller volume. This situation changes as we move to lower apparent magnitudes. As clearly illustrated by Driver et al. (1994), the bright galaxies start to notice the *cosmological volume effect*, i.e. the volume decreases, especially in a closed geometry, whereas the contribution from faint galaxies situated at lower-z still increases with a Euclidean slope. Therefore at faint apparent magnitudes the counts start to be dominated by the intrinsically faint galaxies.

In order to reconcile the predictions from the $q_0 = 0.5$ model with the observations we could introduce a population of intrinsically faint objects (eg. dwarf galaxies, low surface brightness galaxies...) in the LF. In fact we know that dwarfs are very common in our local neighbourhood, e.g. in local clusters galaxies the percentage of dwarfs is as large as ∼ 30 − 50% (Binggeli, Sandage & Tamman 1985; Phillips et al. 1987; Davies et al. 1988).

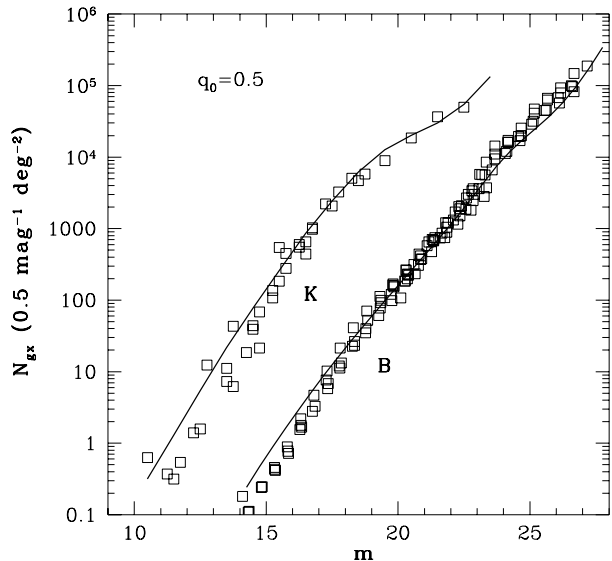

**Figure 6.** The number counts in the B and K bands, and predictions from a $q_0 = 0.5$ luminosity evolution (with dust) model. The faint end slope of late type galaxies is assumed to evolve as $\alpha = \alpha_{z=0} + n * z$ with $n = 0.5$.

If we steepen the faint end slope of the LF in order to fit the counts with the $q_0 = 0.5$ model then we would also predict a large population of low-z objects in the $N(z)$ distributions which is not observed. A possibility is to assume that dwarfs were more numerous in the past, i.e. their number density increases toward higher redshifts. In Figure 6 we plot the number counts in the B and K bands together with predictions from the $q_0 = 0.5$ luminosity evolution model (with dust). The difference between this model and that presented in section 3 is that here we assume that the faint end slope of the late type galaxies LF evolves with the redshift as $\alpha = \alpha_{z=0} + n * z$ with $n = 0.5$. The predictions are now much closer to the observations. On the other hand, the $N(z)$ predictions of this model are quite similar to those shown in section 4, as the contribution of dwarfs starts to be noticeable at the very faint end, beyond the limits of the deep redshift surveys. Nevertheless it is worth mentioning that the models shown in Figure 5 for the 23 <B< 24 sample slightly overpredict the number of galaxies at low-z ($z \sim 0.1 − 0.3$). We suggest that a small percentage of the unidentified galaxies might be low-z dwarfs, where the star formation has recently been stopped and therefore they are still blue but showing featureless spectra (Campos 1995).

The evolution of the faint end slope of the late type galaxies is a rather *ad hoc* ingredient which is based on the models to explain the nature of dwarf galaxies. Dwarf galaxies are thought to form stars in intermittent *bursts*. After a *burst* of star formation is switched on, the small mass of the galaxies might not be able to retain the gas supply against the formation of the galactic wind that might follow the overlap of supernova explosions (Dekel & Silk 1986). If the star formation is abruptly stopped, then the luminosity of the small galaxies will fade away. Therefore if we include the population of dwarfs in the late-type group of galaxies we must allow for luminosity-dependent luminosity evolution, which we have tried to *mimic* by changing the slope of the LF.

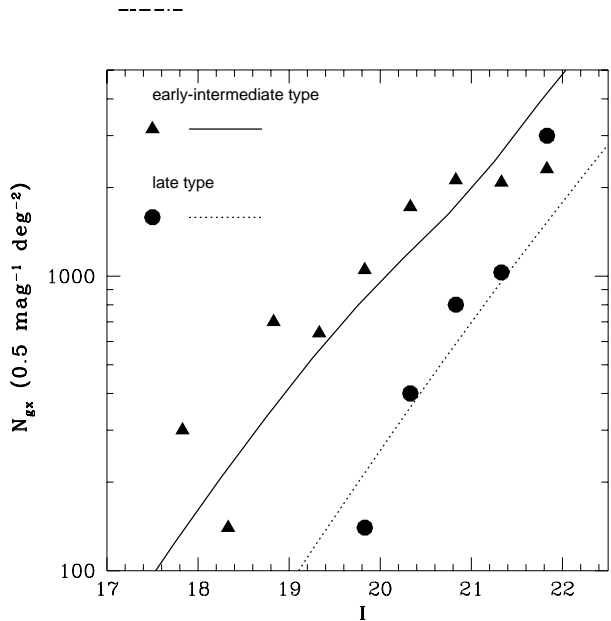

**Figure 7.** The I-band counts of Glazebrook et al. (1995) from the HST Medium Deep Survey, for ellipticals plus spirals (triangles) and irregulars (circles). We also show predictions from the $q_0 = 0.5$ model with evolving faint-end slope.

Using the Hubble Space Telescope HST we are now able to obtain images of very high resolution, such that we can *resolve* the morphology of the faint field galaxies. Therefore it is possible to obtain the number counts for galaxies divided according to their morphology. Recent work on number counts based on HST images (Glazebrook et al., 1995; Driver et al. 1995) has shown that the observed late-type/Irregular population is as much as a factor of $\sim 10$ in excess of the conventional no-evolution models. This excess has been interpreted by these authors as due to a population of dwarfs that has either faded away, or has merged. In Figure 7 we plot the HST counts for early-intermediate galaxies and late/Irregulars (data taken from Glazebrook et al.) together with predictions from the $q_0 = 0.5$ model with evolving LF slope. The model fits the data quite well, except for the last bin where it overpredicts the number of early/intermediate galaxies while underpredicting the number of Irregulars. We regard this discrepancy as possible misidentification of the high-$z$ progenitors of present day ellipticals/spirals or just to the simplicity of the model.

## 6   CONCLUSION

Number counts is a conceptually simple method of measuring the value of $q_0$. However before we can properly address this *cosmological* problem we have to solve an *astrophysical* problem concerning the formation and evolution of the galaxy population. This is proving to be a difficult task, as many different processes, such as evolution of the galaxies luminosity or mergers (see e.g. Guiderdoni & Rocca-Volmerange 1991) are involved.

Even if the modelling might seem a rather complicated problem, we have shown that simple models that consider the evolution of the galaxies luminosity and the extinction produced by dust are able to provide reasonable fits to both the number counts and the redshift distributions of faint field galaxies.

These models only provide a fit to the counts at the faint end when $q_0 = 0.05$. For $q_0 = 0.5$ the models underpredict the counts. A possibility of reconciling models and observations is by means of a population of intrinsically faint objects which becomes increasingly more important at higher redshifts.

99